\documentclass[prl,twocolumn,showpacs,preprintnumbers,amsmath,amssymb,superscriptaddress]{revtex4}

\usepackage{graphicx}
\usepackage{dcolumn}
\usepackage{bm}
\usepackage{color}
\definecolor{red}{rgb}{0.9,0.0,0.1}
\definecolor{green}{rgb}{0.,0.7,0.3}
\input{Qcircuit}
\renewcommand{\ss}{$\sigma_y\otimes\sigma_y$ }

\begin{document}

\title{Quantum Non-Demolition Test of Bipartite Complementarity}

\author{F. de Melo}
\affiliation{Instituto de F\'{\i}sica,
Universidade Federal do Rio de Janeiro, \\Caixa Postal 68528, Rio de
Janeiro, RJ 21941-972, Brazil}
\author{S. P. Walborn}
\affiliation{Instituto de F\'{\i}sica, Universidade Federal do Rio
de Janeiro, \\Caixa Postal 68528, Rio de Janeiro, RJ 21941-972,
Brazil}
\author{J\'anos A. Bergou}
\affiliation{Department of Physics and Astronomy, Hunter College of
the City University of New York, 695 Park Avenue, New York, NY
10021, USA}
\author{L. Davidovich}
\affiliation{Instituto de F\'{\i}sica, Universidade Federal do Rio
de Janeiro, \\Caixa Postal 68528, Rio de Janeiro, RJ 21941-972,
Brazil}

\date{\today}
\begin{abstract}
We present a quantum circuit that implements a non-demolition
measurement of complementary single- and bi-partite properties of a
two-qubit system: entanglement and single-partite visibility and
predictability. The system must be in a pure state with real
coefficients in the computational basis, which allows a direct
operational interpretation of those properties. The circuit can be
realized in many systems of interest to quantum information.
\end{abstract}

 \maketitle
Quantum measurements frequently lead to a ``back-action'' on the
observable being measured. This is the case for instance in the
measurement of the position of a particle, which disturbs its
momentum, thus affecting the future values of its position. This
back-action can be overcome by using a quantum non-demolition (QND)
scheme, introduced in \cite{braginsky77}. In QND measurements, the
observable ${\cal O}_S$ of a system $S$ is measured by detecting a
change in an observable ${\cal O}_P$ of a probe $P$ coupled to $S$
during a finite time, without perturbing the subsequent evolution of
${\cal O}_S$; after the measurement, the final state remains an
eigenvector of ${\cal O}_S$. Experimental implementations have been
performed in the optical domain, for measuring the intensity of an
electromagnetic field \cite{levenson} or the polarization of a
photon~\cite{pryde:190402}, and in cavity quantum electrodynamics
\cite{nogues}. The characterization of QND measurements on qubit
systems were discussed in Ref.~\cite{ralph:012113}.

Extension of the QND concept to bipartite systems poses quite a
challenge, since entanglement measures, like the concurrence
introduced by Wootters~\cite{wootters:2245}, do not have a direct
operational meaning. Also, in the same way that the measurement of
the photon number leads to complete uncertainty on the phase of the
field \cite{brune92}, determination of the entanglement of a pair of
qubits should lead to uncertainty on a complementary variable, and
vice-versa. Identification of this complementary quantity is thus an
important ingredient in understanding the QND scheme.

In this paper, we propose a quantum circuit for QND measurements of
complementary variables in two-qubit systems described by pure
states with real coefficients in the computational basis -- named
rebits in Ref.~\cite{caves:2000}. This restriction allows one to
attach an operational meaning to those variables. One of them is the
concurrence, while the other is a measure of the single-partite
character of the global system. QND determination of the
entanglement of a pair of qubits generates maximally entangled
states even if the incoming state is a product state, analogous to
the QND measurement of the number of photons in a cavity, which
results in a Fock state \cite{brune92}. It also leads to complete
loss of single-partite properties. These are expressed as a sum of
two contributions, standing for predictability and visibility, which
in a double-slit Young interference correspond to the well-known
duality between which-way information and the appearance of
interference fringes. Bi-partite and single-partite properties are
thus seen as complementary aspects, thus generalizing to bi-partite
systems the concept of wave-particle duality.

The concept of complementarity is commonly related to mutually
exclusive properties of single-partite quantum systems, the best
known example being provided by the quantum interpretation of
Young's double-slit experiment. Quantitative relations between
visibility of interference fringes and distinguishability,
corresponding to which-path information, were derived in
\cite{mandel, englert}. Quantification of the concept of
complementarity for multipartite systems is a relatively recent
undertaking. A complementarity relation between single- and
two-particle visibility (which is an intrinsic bipartite property)
was introduced in Refs. \cite{jaeger1}. In \cite{englert2} a
possible connection between the distinguishability and an
entanglement measure was hinted at, and in \cite{jaeger1,abouraddy}
an intimate relation was established between concurrence
\cite{wootters:2245} and the two-particle visibility in an
interferometric setup.  Prompted by these observations, in
\cite{jakob-2003} it was shown that there is an underlying
generalized complementarity relation of which these more restricted
relations emerge as special cases. For two-qubit pure states, the
general complementarity relation of \cite{jakob-2003} can be
mathematically expressed as:
\begin{equation}
{\cal C}^2+{\cal V}_k^2+{\cal P}_k^2=1 \ , \label{triality}
\end{equation}
where the ingredients are the concurrence ${\cal C}$, a genuine
bipartite entity, and the single-partite visibility ${\cal V}_{k}$
and predictability ${\cal P}_{k}$ (for particle $k=1,2$). For mixed
states, the sum of the three terms on the left-hand side of the
above equation is smaller than one.

For a pure state $\ket{\chi}$, the visibility ${\cal V}_k$, a measure of the single-particle coherence (wave-like aspect), is defined as
\begin{eqnarray}
{\cal V}_k =2|\bra{\chi}\sigma_k^+\ket{\chi}|,&\mbox{with
}\sigma_k^+=\left(\begin{array}{cc}
0&1\\
0&0
\end{array}\right) \ .
\label{visi}
\end{eqnarray}
Perfect visibility (${\cal V}_k=1$) is obtained for the states
$(\ket{0}\pm\exp(i\phi)\ket{1})/\sqrt{2}$.

The predictability ${\cal P}_k$, the particle-like aspect, a measure
of the single-particle relative population, is defined for a pure
state as:
\begin{eqnarray}
{\cal P}_k =|\bra{\chi}\sigma_k^z\ket{\chi}|,&\mbox{with
}\sigma_k^z=\left(\begin{array}{cc}
1&0\\
0&-1
\end{array}\right) \ .
\label{predi}
\end{eqnarray}
Perfect predictability (${\cal P}_k=1$) is obtained for the
eigenstates of $\sigma_z$, that is, the states $\ket{0}$ and
$\ket{1}$.

Finally, for pure states the concurrence is defined
as~\cite{wootters:2245}
\begin{eqnarray}
{\cal C}=|\bra{\chi^*}\sigma_y\otimes\sigma_y\ket{\chi}|,&\mbox{with
}\sigma_y=\left(\begin{array}{cc}
0&-i\\
i&0
\end{array}\right)  \ .
\label{concu}
\end{eqnarray}

One should note that the visibility and the predictability are not
invariant under local (single-particle) unitary transformations,
which can actually transform one into the other. For instance, a
unitary transformation takes the maximum-predictability state
$|0\rangle$ into the maximum-visibility state
$(|0\rangle+e^{i\phi}|1\rangle)/\sqrt{2}$. However, the quantity
${\cal{S}}_{k}^{2}={\cal{V}}_{k}^{2}+{\cal{P}}_{k}^{2}$ is invariant
under local unitary transformations, and can be considered as the
proper measure for the single-partitedness of the global system.
With this definition, one can read Eq.\ (\ref{triality}) as a {\it
duality} relation between bipartite and single partite properties,
\begin{equation}
{\cal{C}}^{2}+{\cal{S}}_{k}^{2} =1 \ . \label{bicomp}
\end{equation}
One may say therefore that the single-partite property and the
bipartite property of a two-particle state are complementary just as
the wave and particle properties of single-particle systems are
complementary. While visibility and predictability are properties of
an individual particle, and exhaust for a single-particle system the
full content of wave-particle duality, for a bipartite system the
concurrence, a genuine bipartite quantity, also enters into the
complementarity relation.

We now discuss in detail our method for implementing a QND
measurement of the complementary quantities in Eq.~(\ref{triality}).
To this end we first note that a general bipartite pure qubit state can be
written in the Bell basis as
\begin{equation}
\ket{\chi}=\alpha\ket{\psi^-}+\beta\ket{\psi^+}+\gamma\ket{\phi^-}+\eta\ket{\phi^+}
\ , \label{chi}
\end{equation}
where $\ket{\psi^\pm}=(\ket{10}\pm\ket{01})/\sqrt{2}$,
$\ket{\phi^\pm}=(\ket{11}\pm\ket{00})/\sqrt{2}$ are the Bell states
and  $|\alpha|^2+|\beta|^2+|\gamma|^2+|\eta|^2=1$.

For this state one has:
\begin{eqnarray}
{\cal
V}_{\stackrel{\scriptscriptstyle{1}}{\scriptscriptstyle{2}}}&=&2|\Re(\beta^*\eta
\mp \alpha^* \gamma) + i \Im(\beta \gamma^* \pm\eta\alpha^*)|;\\
{\cal
P}_{\stackrel{\scriptscriptstyle{1}}{\scriptscriptstyle{2}}}&=&2|\Re(\alpha^*
\beta \pm \eta^* \gamma)|;\\
{\cal C}&=&|\alpha^2-\beta^2-\gamma^2+\eta^2|.
\label{concuchi}
\end{eqnarray}

The definition of concurrence involves state conjugation, a
non-physical operation, and therefore this quantity cannot be
directly measured in the general case. For pure states, direct
detection of entanglement has been demonstrated by making a
measurement on two copies of a state~\cite{walborn}. If one measures
just one copy at a time, however, one must further specialize the
state in order for concurrence to be given an operational meaning.
Equation (\ref{concu}) implies  that concurrence is the
magnitude of the average of \ss for all states with real
coefficients in the computational or Bell-state basis. Therefore, it can be
given an operational meaning for this class of states, thus
providing the possibility of directly measuring each term in
Eq.~(\ref{triality}). For this reason, from now on we will be
dealing only with the case of real coefficients. Even though this
limits the general applicability of the method, one should note that
real quantum computation has the full quantum computation power as
was shown in Ref.~\cite{bernstein}.

For this class of states, the visibility is given by
$|\langle\chi|\sigma_x|\chi\rangle|$. Therefore, the quantities in
Eq.~(\ref{triality}) can be expressed in terms of averages of the
operators (taking $k=1$ for definiteness) $\hat
V_1=\sigma_x\otimes\openone$, $\hat P_1=\sigma_z\otimes\openone$,
and $\hat C=\sigma_y\otimes\sigma_y$.  Since these operators do
not commute, a QND measurement of one of them would necessarily
spoil the determination of the other. Thus, for instance, the QND
measurement of $\hat C$ leads to an eigenstate of this observable,
with eigenvalue $\pm1$, yielding a state with maximal concurrence
(equal to one), which is not an eigenstate of $\hat V_1$ or $\hat
P_1$. In fact, the averages of these operators in the resulting
state are equal to zero, thus yielding zero visibility and
predictability, as expected from Eq.~(\ref{triality}).
The uncertainty relation among these three observables is
better expressed in terms of the sum of variances: $(\Delta \hat
V_1)^2+(\Delta \hat P_1)^2+(\Delta \hat C)^2=2$, where
$(\Delta{\cal O})^2=\langle{\cal O}^2\rangle-\langle{\cal
O}\rangle^2$. Since each variance is at most one, this
relation shows that when one of the observables is perfectly
known, the two others must have maximum variance. The uncertainty
relations involving products of variances are not useful in this
case, since for instance $\Delta \hat V_1 \Delta \hat P_1=|\langle
\sigma_y\otimes\openone\rangle|/4$, and the right-hand side
vanishes for an eigenstate of $\hat P_1$, so that $\Delta \hat
V_1$ is undetermined.

We show now that there is a general circuit, involving three
adjustable parameters, which implements QND measurements of these
three observables. We start with a simpler scheme that measures the concurrence, and then consider a more general scheme, which performs a QND measurement of all three quantities
in Eq.~(\ref{triality}).

\begin{figure}[t]
\centering{ \Qcircuit @C=1.5em @R=1.0em { &&
\dstick{\ket{\chi}}&&\gate{R_x(\pi/2)}&\ctrl{2}
& \qw &\gate{R_x(-\pi/2)}& \qw&\qw\\
&&&&\gate{R_x(\pi/2)}&\qw& \ctrl{1} & \gate{R_x(-\pi/2)}&\qw &\qw\\
&&& \lstick{\ket{0}}&\qw&\targ & \targ & \qw& \qw&\meter \\
}} \caption{\footnotesize Quantum circuit for QND measurement of
concurrence. $\ket{\chi}$ is the two-qubit input state, the ancilla
qubit is initially in the state $\ket{0}$ state, and
$R_x(\pi/2)=\exp(-i\pi\sigma_x/4)$.} \label{circuit}
\end{figure}
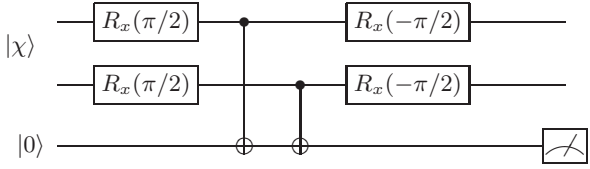

{\em QND measurement of concurrence.} The corresponding circuit is shown in Fig.~\ref{circuit}. It consists of single-qubit rotations and controlled-not (CNOT)
gates, which are the fundamental building blocks of QND measurements.

The composite state, given initially by Eq.~(\ref{chi}) evolves as
follows. The gates $R_x(\pi/2)$ transform it into:
\begin{equation}
\ket{\chi}\ket{0}\rightarrow (\alpha\ket{\psi^-}
-i\eta\ket{\psi^+}+\gamma\ket{\phi^-}-i\beta\ket{\phi^+})\ket{0} \ .
\end{equation}

The two CNOT gates lead to the state:
\begin{equation}
(\alpha\ket{\psi^-}-i\eta\ket{\psi^+})\ket{1}+(\gamma\ket{\phi^-}
-i\beta\ket{\phi^+})\ket{0} \ .
\end{equation}

After the final rotations one has:
\begin{equation}
(\alpha\ket{\psi^-}+\eta\ket{\phi^+})\ket{1}+(\gamma\ket{\phi^-}
+\beta\ket{\psi^+})\ket{0} \ .
\end{equation}
The final step is to perform the measurement on the ancilla state.
Thus the conditional outgoing states are:
\begin{equation}
\left\{\begin{array}{cc}
    \ket{\chi_1}=\frac{\alpha\ket{\psi^-}+\eta\ket{\phi^+}}{\sqrt{\alpha^2+\eta^2}}&\mbox{if
the ancilla is in}\;\ket{1}\\
\\
  \ket{\chi_0}=\frac{\gamma\ket{\phi^-}+\beta\ket{\psi^+}}{\sqrt{\beta^2+\gamma^2}}&\mbox{if
the ancilla is in}\;\ket{0}
       \end{array}\right. .
\label{states}
\end{equation}
The concurrence of these states is easily calculated to be:
\begin{equation}
\left\{\begin{array}{ccl}
    {\cal C}(\chi_1)&=&{|\alpha^2+\eta^2|}/{(\alpha^2+\eta^2)}=1\\
    {\cal C}(\chi_0)&=&{|-\beta^2-\gamma^2|}/{(\beta^2+\gamma^2)}=1
       \end{array}\right. .
\label{concu10}
\end{equation}

The concurrence of the outgoing state is equal to $1$ independent of
the result of the ancilla measurement, provided the coefficients in
the initial state are real. Thus, the outgoing state is maximally
entangled for any input state.  We can even start with a separable
state such as $\ket{00}$, for example, and the final state will
still be either $\ket{\phi^+}$ or $\ket{\phi^-}$, depending on the
ancilla measurement outcome. The concurrence of the initial state is
determined from the statistics of the measurements on the ancilla
for many realizations of the experiment:
\begin{equation}
|p_{1}-p_{0}|=|\alpha^2-\beta^2-\gamma^2+\eta^2|= C(\chi) \,;
\end{equation}
$p_i$ being the probability of finding the ancilla in state $i$.

Although entanglement is invariant under local transformations, we undo the rotations in order to end up in a \ss eigenvector, avoiding then the back action. The circuit
thus measures in a QND way the expectation value
$\langle\sigma_y\otimes\sigma_y\rangle$, the magnitude of which is
the concurrence for real states.

\begin{figure}[t]
\centering{ \Qcircuit @C=1.5em @R=1.0em {
&& \dstick{\ket{\chi}}&&\gate{R_{\vec{\theta_1}}}&\qw&\ctrl{2} & \qw & \gate{R_{\vec{\theta_2}}} &\qw&\qw\\
&&&&\gate{R_{\vec{\theta_1}}}&\qw &\qw& \ctrl{2} & \gate{R_{\vec{\theta_2}}} & \qw &\qw\\
&&& \lstick{\ket{0}}&\gate{R_{\vec{\theta_3}}}&\ctrl{1}&\targ & \qw & \qw& \qw &\meter \\
&&&\lstick{\ket{0}}&\qw&\targ&\qw & \targ & \qw& \qw &\meter
\gategroup{3}{5}{4}{6}{1em}{--} }} \caption{\footnotesize Universal
quantum circuit for QND measurement of concurrence, visibility and
predictability. The dashed box is the ancilla state preparation and
$R_{\vec\theta_i}=\exp(-i\vec\sigma\cdot\vec\theta_i)$.}
\label{circuit2}
\end{figure}
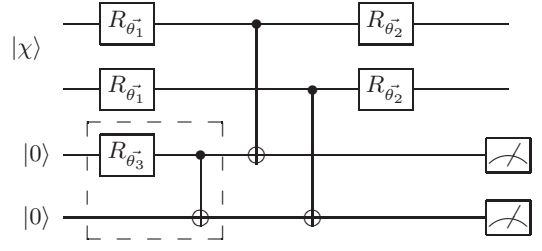

{\em QND measurement of single- and bi-partite features.} In order
to perform QND measurements of all the observables corresponding
to the quantities in the complementarity relation~(\ref{triality}),
we need a circuit that allows the measurement of single-particle
features as well. Such a circuit is presented in
Fig.~\ref{circuit2}.

The previous result can be obtained by setting
$\vec{\theta_3}=(\pi/2)\hat{y}$, so that the ancilla is prepared
in a maximally entangled state $\ket{\phi^+}$. Only $\ket{\phi^+}$
and $\ket{\psi^+}$ are used, and these states act as a logical
qubit. This circuit is then completely equivalent to the one in
Fig.~\ref{circuit}, replacing $\ket{0}\rightarrow\ket{\phi^+}$ and
$\ket{1}\rightarrow\ket{\psi^+}$. Here
$\vec\theta_1=(\pi/2)\hat{x}=-\vec\theta_2$, as in the previous circuit. A
local measurement in the computational basis distinguishes between $\ket{\phi^+}$ and
$\ket{\psi^+}$. The probabilities are easily related:
$p_{\phi^+}=p_{00}+p_{11}$ and $p_{\psi^+}=p_{10}+p_{01}$. The
concurrence is now given by $|p_{\psi^+}-p_{\phi^+}|$.

For the QND measurement of $\hat V_k$ and $\hat P_k$, corresponding
to visibility and predictability, we choose $\vec{\theta_3}=0$,
which leads to a separable ancilla state $\ket{00}$. In this case
the odd lines of the circuit are completely decoupled from the even
ones, thus yielding two independent circuits, which is a natural
choice if one wants to measure single-particle aspects.

A single CNOT gate, without any rotation, would project the final
state onto an eigenvector of $\sigma_z$, that is, onto one of the
computational-basis states. The average of the measurements on the
ancilla yield $\langle \sigma_z \rangle$, and hence this is a
non-demolition measurement of the predictability. Therefore,
measurement of the predictability corresponds to the choice
$\vec\theta_1=\vec\theta_2=0$.

The state before the measurement of the ancilla is then:
\begin{eqnarray}
\frac{1}{\sqrt{2}}[(\eta-\gamma)\ket{00}\ket{00}+(\beta-\alpha)\ket{01}\ket{01}+\nonumber\\
(\alpha+\beta)\ket{10}\ket{10}+(\gamma+\eta)\ket{11}\ket{11}].
\end{eqnarray}

A measurement on the ancilla leads to an outgoing state with perfect
predictability for both qubits. The probabilities for the several
possible outcomes yield the predictabilities of the initial real state:
\begin{equation}
\left\{\begin{array}{c}
        |(p_{00}+p_{01})-(p_{10}+p_{11})|= 2|\alpha\beta+\eta\gamma|={\cal P}_1\\
        |(p_{00}+p_{10})-(p_{01}+p_{11})|= 2|\alpha\beta-\eta\gamma|={\cal
        P}_2
        \end{array}\right..
       \end{equation}
The first line represents the difference between the probabilities
of measuring $0$ and $1$ for the first ancilla, while the second
line is the difference between the probabilities of finding the
second ancilla in either $0$ or $1$.

For the QND measurement of the visibility, one must perform a
$\pi/2$ rotation around the $\hat{y}$ axis in state space, since the
visibility for real states is related to the $\sigma_x$ matrix.
However, the visibility does change under local rotations, therefore
the initial rotation must be undone at the end of the circuit, in
order to end up in a maximum visibility state for both qubits. Thus,
one must have $\vec\theta_1=(\pi/2)\hat{y}=-\vec\theta_2$, which
leads to the following composite state right before the ancilla
measurement:
\begin{eqnarray}
\frac{1}{\sqrt{2}}[(\eta+\beta)\ket{+}\ket{+}\ket{00}+
(\gamma-\alpha)\ket{+}\ket{-}\ket{01}&+\nonumber\\
(\alpha+\gamma)\ket{-}\ket{+}\ket{10}+(\eta-\beta)\ket{-}\ket{-}\ket{11}]&,
\end{eqnarray}
where $\ket{\pm}\equiv(\ket{1}\pm \ket{0})/\sqrt{2}$.

The ancilla measurement in the computational basis will project the outgoing state onto a maximum
visibility state. From the measurement statistics one can infer the
initial-state visibility for both qubits:
\begin{eqnarray}
\left\{\begin{array}{c}
        |(p_{00}+p_{01})-(p_{10}+p_{11})|= 2|\alpha\gamma-\eta\beta|={\cal V}_1\\
        |(p_{00}+p_{10})-(p_{01}+p_{11})|= 2|\alpha\gamma+\eta\beta|={\cal V}_2
       \end{array}\right. \ .
\end{eqnarray}
With these two measurements one has a full QND characterization of the single-particle features. The outgoing state in both cases is separable.

It is easy to check that the above measurement scheme fulfills all
the requirements for qubit QND measurements listed in
Ref.~\cite{ralph:012113}. The outgoing state is, after measurement,
an eigenvector of the measured observable. For instance, when
measuring concurrence the state of the system becomes a \ss
eigenvector. Also, ${\cal{S}}_{k}^{2}$ and the concurrence do not
change in time due to free local evolution. The requirements for QND
measurements are then fulfilled for both parts of the
complementarity relation in Eq.~(\ref{bicomp}). On the other hand,
visibility and predictability can be interchanged between each other
depending on the free Hamiltonian. For many cases of interest,
however, the free Hamiltonian is proportional to
$(\sigma_z\otimes\openone+\openone\otimes\sigma_z)$ and, in these
cases, the visibility and predictability measurements are themselves
QND-like.

The above circuits can be implemented in many systems of interest
for quantum information, since they involve single-particle
rotations and CNOT gates, which have been demonstrated for instance in trapped
ions~\cite{kaler}, cavity QED~\cite{haroche} and with two pairs of twin
photons, created as shown in Ref.~\cite{zhao}.

In conclusion, we have shown that it is possible to implement
independent QND measurements of all the complementary quantities
corresponding to a two-qubit state, which express its single- and
bipartite content. The restriction to states with real coefficients
in the computational basis seems to be unavoidable in the present
context, since otherwise it is not possible to attribute an
operational meaning to concurrence for measurements realized on
single copies of an ensemble.

These measurements illustrate the complementarity among single- and
bipartite quantities: a QND measurement of entanglement leads to a
maximally entangled state, but spoils at the same time the
visibility and the predictability for each qubit. This could have
broad implications for quantum information processing, since after
determining the single-partite or bi-partite content of a quantum
state, the state itself can be further processed; elimination of
back-action guarantees that the measured value is preserved. Thus,
after a measurement of entanglement, the resulting state could be
used as a resource in, e.g., teleportation~\cite{bennett:1895} and
quantum cryptography~\cite{bennett:3121} protocols.

\begin{acknowledgments}
We would like to thank L. Aolita and R.L. de Matos Filho for useful
discussions. JB is grateful to the Institute of Physics of the
Federal University of Rio de Janeiro for the hospitality extended to
him during his stay there.  This work was supported by the Brazilian
agencies CAPES, CNPq, FAPERJ, FUJB, and the Millennium Institute for
Quantum Information.
\end{acknowledgments}


\begin{thebibliography}{10}

\bibitem{braginsky77}
V.~Braginsky, Y.~Vorontsov, and F.~Khalili,
\newblock Zh. Eks. Teor. Fiz. {\bf 73}, 1340 (1977).

\bibitem{levenson}
M.~Levenson {\em et~al.},
\newblock Phys. Rev. Lett. {\bf 57}, 2473 (1986). N.~Imoto, S.~Watkins, and Y.~Sasaki, \newblock \oc {\bf 61}, 159 (1987). A.~LaPorta, R.E.~Slusher, and B.~Yurke, \newblock Phys. Rev. Lett. {\bf 62}, 28 (1989).

\bibitem{pryde:190402}
G.~J. Pryde {\em et~al.},
\newblock Phys. Rev. Lett. {\bf 92}, 190402 (2004).

\bibitem{nogues}
G.~Nogues {\em et~al.},
\newblock Nature {\bf 400}, 239 (1999).

\bibitem{ralph:012113}
T.~C. Ralph {\em et~al.},
\newblock Phys. Rev. A {\bf 73}, 012113 (2006).

\bibitem{wootters:2245}
W.~K. Wootters,
\newblock Phys. Rev. Lett. {\bf 80}, 2245 (1998).

\bibitem{brune92}
M.~Brune {\em et~al.},
\newblock Phys. Rev. A {\bf 45}, 5193 (1992).

\bibitem{caves:2000}
C.~M. Caves, C.~A. Fuchs, and P.~Rungta,
\newblock Found. Phys. Lett. {\bf 14}, 199 (2001).

\bibitem{mandel}
L. Mandel,
\newblock Opt. Lett. \bf 16\rm, 1882 (1991).

\bibitem{englert}
B.-G. Englert,
\newblock Phys. Rev. Lett. {\bf 77}, 2154 (1996).

\bibitem{jaeger1}
G.~Jaeger, M.A.~Horne, and A.~Shimony,
\newblock Phys. Rev. A {\bf 48}, 1023 (1993). G.~Jaeger, A.~Shimony, and L.~Vaidman, \newblock Phys. Rev. A {\bf 51}, 54 (1995).

\bibitem{englert2}
B.-G. Englert and J.~Bergou,
\newblock \oc {\bf 179}, 337 (2000).

\bibitem{abouraddy}
A.~F. Abouraddy {\em et~al.},
\newblock Phys. Rev. A {\bf 64}, 050101(R) (2001). M.~Jakob and J.~Bergou,
\newblock Phys. Rev. A {\bf 66}, 062107 (2002).

\bibitem{jakob-2003}
M.~Jakob and J.~A. Bergou,
\newblock quant-ph/0302075 (2003).

\bibitem{walborn}
S.P.~Walborn {\em et~al.},
\newblock Nature {\bf 440}, 1022 (2006).

\bibitem{bernstein}
E.~Bernstein and U.~Vazirani,
\newblock Siam J. of Comp. {\bf 26}, 1411 (1997).
Y.~Shi,
\newblock quant-ph/0205115.

\bibitem{kaler} F. Schmidt-Kaler {\em et~al.}, Nature {\bf 422}, 408
(2003); D. Leibfried {\em et~al.}, Nature {\bf 422}, 412 (2003).

\bibitem{haroche}
A. Rauschenbeutel {\em et~al.}, Phys. Rev. Lett. \bf 83\rm, 005166
(1999).

\bibitem{zhao}
S. Gasparoni {\em et~al.} \newblock Phys. Rev. Lett. {\bf 93}, 020504 (2004).

\bibitem{bennett:1895}
C.~H. Bennett {\em et~al.},
\newblock Phys. Rev. Lett. {\bf 70}, 1895 (1993). J.-W. Pan {\em et~al.},
\newblock Nature {\bf 421}, 721 (2003).

\bibitem{bennett:3121}
C.~H. Bennett,
\newblock Phys. Rev. Lett. {\bf 68}, 3121 (1992).


\end{thebibliography}

\end{document}